# Usage des Intelligences artificielles dans la création musicale : entre interface et appropriation

*The use of artificial intelligence in music creation: between interface and appropriation*

< Arnaud ZELLER [1] > < Emmanuelle CHEVRY PEBAYLE [2] >

1. Université de Strasbourg
   zellera@unistra.fr

2. Université de Strasbourg
   chevry@unistra.fr

< RÉSUMÉ >

En observant les activités et les relations des musiciens et designers sonores aux activités de création, d'interprétation, d'édition et de diffusion avec l'intelligence artificielle (IA), depuis deux forums spécialisés entre 2022 et 2024, cet article propose une analyse des représentations liées à leur utilisation. En effet, la machine, désormais dotée d'intelligences artificielles nécessitant de nouvelles appropriations et permettant de nouvelles médiations, constitue de nouveaux enjeux pour les artistes. Pour étudier ces confrontations et ces nouvelles médiations, notre approche mobilise le cadre théorique de l'Human-AI Musicking Framework, à partir d'une analyse lexicométrique de contenus. Il s'agit d'expliciter les usages présents et à venir de l'IA depuis les interfaces, dans le cadre de création de contenus sonores et musicaux et de relever les obstacles, freins et limites constatés à l'appropriation « dans le fait de faire sien le contenu et de l'intégrer comme une part de soi » (Bachimont et Crozat, 2004) dans le cadre d'une collaboration entre le musicien et la machine.

< MOTS-CLÉS >

intelligences artificielles, interface, médiations, appropriation, collaboration créative

**< ABSTRACT >**

By observing the activities and relationships of musicians and sound designers to the activities of creation, performance, publishing and dissemination with artificial intelligence (AI), from two specialized forums between 2022 and 2024, this article proposes a lexicometric analysis of the representations linked to their use. Indeed, the machine, now equipped with artificial intelligences requiring new appropriations and enabling new mediations, constitutes new challenges for artists. To study these confrontations and new mediations, our approach mobilizes the theoretical framework of the Human-AI Musicking Framework, based on a lexicometric analysis of content. The aim is to clarify the present and future uses of AI from the interfaces, in the creation of sound and musical content, and to identify the obstacles, obstacles, brakes and limits to appropriation "in the fact of making the content one's own and integrating it as a part of oneself" (Bachimont and Crozat, 2004) in the context of a collaboration between musician and machine.

**< KEYWORDS >**

artificial intelligence, interfaces, mediations, appropriation, creative collaboration

**Introduction**

L'essor de l'intelligence artificielle transforme aujourd'hui l'ensemble de l'activité humaine dans des secteurs aussi éloignés que la santé, la finance, les transports ou l'éducation (Russell et Norvig, 2021). Le domaine de la créativité définie comme ce qui requiert de l'originalité et de l'efficacité (Runco *et al.*, 2012), ou encore depuis les processus dynamiques de créativité universelle (DUCP ; Corazza, 2019a) comme un ensemble actif de tous les épisodes qui mettent en jeu une structure arborescente d'épisodes de créativité interconnectés (Corazza, 2020), sont eux aussi touchés par cette révolution (Makridakis, 2017). L'utilisation de technologies informatiques pour créer ou aider à créer de la musique n'est pas récente comme en témoigne Ada Lovelace dans ses notes sur la machine analytique de Babbage dans lesquels elle explique que « la machine pourrait composer des morceaux de musique élaborés et scientifiques de tout degré de complexité ou d'étendue » (Civit *et al.*, 2022). Au-delà des tâches d'apprentissage automatique, le Deep learning peut aussi être utilisé pour l'analyse musicale (Briot, 2021). Les intelligences artificielles (IA) permettent aujourd'hui aux créateurs de développer des mélodies, des harmonies, des paroles et des mélanges en appuyant simplement sur un bouton (Newman, 2023). Si le numérique peut soutenir l'activité humaine, il peut aussi provoquer une confusion entre les supports et des signaux. Les études conduites sur l'application des intelligences artificielles à la musique montrent que celles-ci favorisent le progrès, en permettant de faire évoluer les styles musicaux, par la génération de musique automatique ou en concourrant à la conduite d'études et d'exploration spécifique, et qui conduisent, dans une certaine mesure, au renouvellement de la musique occidentale ou permettent de développer du nouveau matériel et de nouvelles applications pédagogiques en musique (Mora-Gutiérrez *et al.*, 2024). L'un de ces usages les plus courants étant l'automatisation des processus compositionnels, afin d'analyser et de comprendre des structures musicales à partir d'analyses bayésiennes, ou encore de faciliter l'éducation musicale.

Dans cet article, nous proposons d'interroger les (re)compositions de pièces musicales issues des IA mobilisées, en analysant les profils des musiciens utilisateurs de ces intelligences, leurs usages et les buts recherchés. Il s'agit de comprendre, d'une part, les logiques qui orientent les choix des musiciens de mobiliser ces IA dans le cadre de créations, et d'autre part, de recueillir les avis de ces musiciens sur l'évolution de leur rapport à la création depuis leurs perceptions des nouvelles médiations à l'œuvre dans l'utilisation de ces technologies et des résultats qu'elles produisent. Plus spécifiquement, si ces IA peuvent être appréhendées comme des nouvelles déclinaisons des compagnons virtuels déjà présents depuis plusieurs décennies dans le numérique, qu'en est-il des médiations qu'elles génèrent et de l'inspiration qu'elles sont censées soutenir ? L'opérationnalisation de l'idée musicale par le compositeur avec ces IA, mobilise-t-elle les mêmes types de collaboration musicien-machine que l'usage du numérique avec les logiciels d'aide à la composition, du point de vue des interfaces ou des procédés numériques déjà présents dans les jeux permettant de composer des pièces musicales aléatoirement, jeux publiés à partir du XVIII[e] siècle (Lorrain, 2003) et s'appuyant notamment sur l'algorithmie et le hasard des combinaisons ? Comment les musiciens s'approprient-ils les médiations issues des collaborations Humain-IA suite aux reconfigurations des espaces de travail induits par l'usage de ces technologies dans la création ? Quelles évolutions ces médiations induisent-elles de la part des musiciens utilisateurs dans leurs rapports à la composition et aux intelligences artificielles ? Il s'agit donc de questionner les confrontations et les collaborations culturelles entre le musicien et la machine d'une « transition numérique responsable » et d'une numérisation adaptée aux besoins des artisans (Kellner *et al.,* 2010), qu'ils soient organisationnels ou économiques.

Dans le premier volet de notre étude, nous rappelons le cadre théorique de l'appropriation en musique puis rappelons le cadre théorique de l'Human-AI Musicking (Vear *et al*., 2023), méthodologie proposée pour analyser et caractériser la co-créativité entre le musicien et la machine. Dans ce cadre, la musique est appréhendée dans une perspective actancielle, le *musicking*, un savoir-faire de la musique (*Doing music*). Cette méthodologie est appliquée dans le deuxième volet de notre article où nous revenons sur les usages des musiciens utilisateurs d'IA. Il s'agit de présenter les résultats de l'analyse des usages de ces IA en observant notamment leurs motivations pour produire du matériau musical, et l'évaluation qu'ils portent sur les résultats obtenus depuis celles-ci à partir des intelligences mobilisées. Nous cherchons à comprendre l'expérience mais aussi les croyances et les représentations autour de ces intelligences, depuis les récits de deux communautés impliquées dans des activités de production musicale, la communauté issue du forum de discussion KVR et la communauté issue du forum de discussion Gearspace. Dans ce deuxième volet, nous tentons aussi de clarifier, toujours à partir des mêmes sources, ce qui constitue des leviers mais aussi des obstacles, des freins et des limites constatés à l'appropriation, depuis l'interface, de l'intelligence artificielle générative et son introduction dans le geste musical « dans le fait de faire sien le contenu et de l'intégrer comme une part de soi » (Bachimont et Crozat, 2004). Pour ce faire nous mobilisons un sondage en ligne, afin de susciter des réactions et des prises de positions.

Nous proposons enfin dans notre troisième volet, une discussion sur les perspectives de nouvelles collaborations envisagées avec ces intelligences artificielles par les musiciens qui cherchent à adapter leur visibilité et les moyens de se faire reconnaître en réponse à l'évolution de l'industrie musicale, en réponse à des nouvelles pratiques culturelles, de nouvelles formes de consommation musicale, des nouvelles technologies utilisables mais aussi de l'hyper concurrence, compte-tenu de la démocratisation des ressources d'aide à la composition et à la diffusion mises à disposition des créateurs, et des demandes des professionnels de ce secteur.

**Contexte de la recherche**

Le recours à l'intelligence artificielle en musique fait l'objet de nombreuses études. Les technologies qui découlent de l'IA constituent des outils permettant d'accroître la créativité humaine en générant de nouvelles médiations censées favoriser l'expression de nouvelles idées musicales en fournissant de l'inspiration. En effet, des auteurs ont montré que les IA génératives facilitent le processus de composition en produisant rapidement une grande quantité de contenus, permettant de surmonter les blocages créatifs, et qu'elles stimulent le développement ultérieur de morceaux musicaux (Tegmark, 2017 ; Civit *et al.,* 2022). Pour autant d'autres auteurs (Dash et Agres, 2024) ont relevé plusieurs problèmes pour ces systèmes comme le *contrôle*, qui se réfère à la possibilité pour l'utilisateur de spécifier le contenu émotionnel souhaité de la musique générée, et à la capacité du système à contrôler précisément les caractéristiques musicales de manière à ce que la musique résultante présente l'effet désiré. Un autre problème relevé est celui de *l'adaptabilité narrative*, aujourd'hui insuffisante, et qui fait référence à la capacité des ces systèmes à générer un morceau de musique cohérent pouvant s'adapter à l'exigence émotionnelle de la narration. Enfin, parmi les autres défis majeurs relevés, figure *l'hybridation* qui consiste à combiner plusieurs techniques de génération de musique (humaine et artificielle), à partir de médiations, dans un seul ou plusieurs systèmes, avec des structures musicales relativement longues, sans que les structures déjà existantes soient répétées.

**1. Médiation et collaboration**

Pour Jeanneret et Ollivier (2004), le terme médiation désigne […] « l'espace dense des constructions qui sont nécessaires pour que les sujets, engagés dans la communication, déterminent, qualifient, transforment les objets qui les réunissent, et établissent ainsi leurs relations ». Dans le domaine artistique, la médiation est appréhendée comme « ce qui fait l'art, juste entre le geste et la chose » (Hennion, 1993a), dans des espaces et lors d'activités durant lesquelles se produisent des « procédures de médiation » (Hennion, 1993b).

Nous approchons le concept de médiation depuis la collaboration entre l'humain et la machine. À partir des cadres théoriques définis en psychologie sociale et des études conduites sur la créativité issues de la psychologie culturelle (Barrett *et al.*, 2014 ; Glăveanu, 2015 ; Glăveanu *et al.*, 2014), des chercheurs remettent de plus en plus en question les conceptions individualistes de la créativité, affirmant que l'interaction sociale, la communication et la collaboration sont des éléments clés de la pensée et de la pratique créatives (John-Steiner, 2020). Ainsi, des auteurs appréhendent la collaboration comme une caractéristique centrale de la créativité (Schiavio *et al.*, 2019), ou la créativité comme nécessitant d'être intégrée dans la collaboration (Kenny, 2014).

En nous situant dans le cadre de la collaboration créative et de la créativité collaborative (Barrett *et al.*, 2021), nous posons que la créativité musicale avec l'IA, qu'il s'agisse de performance musicale, d'improvisation ou encore de composition réalisée avec des intelligences artificielles génératives, peut être considérée comme un exemple de collaboration créative et/ou de créativité collaborative entre l'humain et la machine, à travers les possibilités (ou contraintes) historico-socio-culturelles qu'elle offre et les éléments pouvant être combinés tels que des matériaux, des technologies, des humains (compositeurs, interprètes, chefs d'orchestre, développeurs d'outils, assistants musicaux, etc.) et des actions incarnées.

### 1.1. Appropriation et IA en musique

L'introduction de l'intelligence artificielle dans l'activité humaine reconfigure les espaces de travail (Ashri, 2019) à l'intérieur desquels les outils et des technologies qui sont utilisés mettent en jeu des collaborations mais aussi des processus d'appropriation, en raison des interactions et technologies utilisées.

Dans la littérature, le concept d'appropriation est examiné sous différentes perspectives en mettant l'accent, selon les disciplines, sur les aspects sociologiques, techniques, psychologiques ou cognitifs. Du point de vue sociologique, l'appropriation est définie comme la personnalisation ou la manière dont la technologie et les artefacts technologiques sont adoptés, façonnés et ensuite utilisés par l'individu (Kirk *et al.*, 2015). Plus précisément, Janneck (2009) définit l'appropriation comme un processus par lequel les utilisateurs intègrent la technologie dans leur vie, leurs pratiques et leurs routines de travail. Du point de vue technique, l'appropriation est appréhendée en tant que modification de l'usage prévu (Beenkens et Verburg, 2007). Une telle adaptation de l'utilisation permet aux participants d'obtenir davantage des TIC que leur objectif initial, bien que cette utilisation innovante n'ait peut-être pas été prévue par les concepteurs (Dey *et al.*, 2013). Du point de vue psychologique, des chercheurs (O'Brien et Toms, 2008) ont montré que l'appropriation requiert un engagement émotionnel et cognitif et que l'engagement d'un individu envers les TIC dépend de la manière dont il développe sa connexion avec les technologies (Garnham, 1999).

À partir du concept d'utilisabilité, Nielsen (1994) définit la facilité d'appropriation comme « la facilité à se souvenir ». L'approche holistique permet de saisir les différents niveaux de l'appropriation : micro (activité et utilisation de la technologie), méso (cadre social de l'activité) et macro (réseau sociotechnique et jeux d'acteurs). En présentant son modèle de fracture numérique, Selwyn (2004) considère l'appropriation comme une « utilisation significative des TIC » où l'utilisateur « exerce un certain degré de contrôle et de choix sur la technologie et son contenu, ce qui donne un sens, une signification et une utilité à l'individu concerné ».

Les processus d'appropriation qui témoignent de la manière dont les utilisateurs se saisissent de leur espace, constituent des enjeux cruciaux dans le cas des espaces de travail par activités (Lai *et al.*, 2024). Articulant les perspectives de l'ergonomie et de la sociologie des usages, ces processus d'appropriation sont généralement appréhendés selon trois niveaux d'analyse possibles : (i) les caractéristiques physiques de l'environnement, comme le confort physique, (ii) son adéquation par rapport aux tâches devant être réalisées (confort fonctionnel), et (iii) les dimensions psychosociales, comme le sentiment de contrôle sur l'espace de travail.

Dans notre étude, nous nous intéressons à ce troisième niveau à savoir l'analyse des processus d'appropriation de l'IA à des fins de création musicale à partir des dimensions psychosociales, puisque, en référence à la genèse instrumentale (Rabardel, 1995), des usages déviants de ces espaces de travail ont été observés par rapport à ce qui a été prévu et anticipé par les concepteurs (Cobaleda et Babapour, 2017 ; Vischer, 2008) ; et que selon Jeanneret et Ollivier (2004), la médiation est une pratique qui « n'est jamais, ni immédiate, ni transparente ».

Dans le domaine de l'éducation musicale, les intelligences artificielles sont considérées comme un des moyens permettant de personnaliser les expériences d'apprentissage à partir d'agents conversationnels intelligents pouvant s'adapter aux besoins individuels des étudiants, en fournissant un enseignement et un retour d'information personnalisés qui améliorent les résultats (Li & Wang, 2022 ; Bamigbola, 2021). À l'université, une étude montre que l'intégration de ces technologies dans les situations pédagogiques informatisées impacte la réussite des étudiants en éducation musicale avec des variations observées, les résultats scolaires des étudiants pouvant être attribuées à l'influence combinée de leur auto-efficacité et de leur préparation à l'IA (Wang et Li, 2024). Dans la recherche, ces intelligences artificielles génératives jouent également un rôle crucial en permettant par exemple la recherche d'informations musicales et l'analyse technique automatisée, avec la possibilité pour les éducateurs et les étudiants de classer et d'évaluer les compositions musicales en fonction de leurs caractéristiques, afin de mieux comprendre la musique (Ben-Tal *et al.*, 2021 ; Gouzouasis & Bakan, 2011).

Les communautés en ligne qui servent de plateformes de collaboration, de partage des connaissances et de promotion de la créativité parmi les musiciens, les compositeurs et les développeurs d'IA, se sont emparées du sujet et des enjeux qui en découlent. Dans les discussions sur les outils d'IA, elles aident les utilisateurs à naviguer dans la complexité liée à l'intégration des outils proposés dans les différents stades ou les différents contextes dans lesquels elles s'intègrent aux pratiques musicales (Chamberlain *et al.*, 2021 ; Blackwell *et al.*, 2021). Ces communautés jouent également un rôle crucial dans la démocratisation de l'accès aux ressources de production musicale et aux technologies d'IA. Elles permettent à des personnes d'horizons divers de se connecter et de partager leurs expériences en apprenant les unes des autres, ce qui conduit à des utilisations innovantes de l'IA dans la création musicale (Ben-Tal *et al.*, 2021 ; Gouzouasis & Bakan, 2011). En outre, ces communautés organisent souvent des événements, des ateliers et des projets collaboratifs qui encouragent la participation et l'engagement, améliorant globalement l'expérience d'apprentissage des membres (Dan *et al.*, 2003 ; Ernst *et al.*, 2019). Parmi les sujets âprement discutés, revient régulièrement la question de la création avec l'intelligence artificielle, d'où l'intérêt d'observer les médiations et ce qui les caractérise à partir de l'analyse des contributions.

### *1.2. Médiations : approche des relations créatives entre l'humain et l'IA avec l'Human-AI Musicking*

Le Human-AI Musicking Framework (Vear *et al.*, 2023) de l'Université de Nottingham est un cadre théorique qui explore les relations créatives entre les musiciens, en action, et l'IA, dans le but d'améliorer la conception des systèmes d'IA dans le domaine de la musique. L'objectif de ce cadre est de fournir une méthode et un outil qui rapprochent progressivement le musicien des relations incarnées qui pourraient former le sens dans la perspective de la musique, à partir des différents moments du processus de création, du temps de la conception, aux temps de prototypage, de tests utilisateurs et de raffinement, jusqu'à l'évaluation. Ce cadre théorique présente une série de questions structurées pour expliciter, (a) comment l'IA créative se manifeste dans la relation entre l'homme et l'IA, (b) comment elle fonctionne, et (c) comment elle interagit avec le musicien. En adoptant une approche ascendante, incarnée et phénoménologique, les auteurs fournissent un outil permettant aux chercheurs de comprendre comment les musiciens peuvent, ou non, s'engager de manière significative avec ces intelligences.

Ce cadre est constitué de cinq éléments. Le premier élément est *l'objectif créatif de l'IA créative*, qui traite de la manière dont les objectifs de l'IA créative se manifestent dans le flux incarné de la musique. Il détermine les objectifs spécifiques de l'IA dans un scénario de musique, comme la génération de mélodies ou le contrôle de paramètres en temps réel basés sur des gestes humains. Le deuxième élément est *l'interaction* qui vise à identifier la manière dont les caractéristiques de l'IA créative s'inscrivent dans la temporalité de la musique, en prenant en compte le type d'interaction que l'IA suscite, qu'elle soit ludique ou pragmatique, et en examinant les modes d'interaction créative, tels que l'interaction simultanée, la collaboration ou la cocréation. Le troisième élément est *le rôle de l'être humain dans la boucle* qui tente d'évaluer la manière dont l'IA créative influence la créativité humaine et dans quelle mesure elle s'aligne sur les besoins et les expériences des musiciens. Le quatrième élément est *l'impact sur les musiciens humains* qui examine la manière dont l'IA peut influencer le rapport des musiciens à la musique.

Le dernier élément est celui des *Relations corporelles* qui permet d'éclairer la compréhension des relations incarnées qui peuvent donner un sens à la musique, en se concentrant sur la manière dont ces relations évoluent tout au long du processus de conception, y compris la conception, le prototypage, les tests utilisateurs, le raffinement et l'évaluation. La collaboration artistique possible entre l'humain et l'IA, dynamique et souvent imprévisible, peut s'avérer productive selon la volonté et de la capacité des utilisateurs (musiciens) à s'engager avec l'IA pour produire des processus et du matériau observable. Même si les auteurs reconnaissent que, de par la nature même des questions posées et des éléments structurés proposés, les résultats pourraient renvoyer à une simplification excessive du processus créatif, ce processus permettrait toutefois de mettre au jour de nouveaux rapports ambivalents et de nouvelles formes d'interactions. Des cadres d'analyse comme le Human-AI Musicking Framework permettent d'ailleurs d'étudier la complexité de ces interactions entre l'homme et l'IA même si certaines caractéristiques ou nuances de ces interactions peuvent parfois être difficiles à appréhender ou à catégoriser, la nature subjective de la musique et des expériences individuelles ne pouvant être ramenée à un ensemble de questions structurées.

### *1.3. Les communautés de musiciens observées*

De nombreuses communautés en ligne de musiciens favorisent la collaboration et le partage des connaissances en ligne, dans le domaine de la musique et de l'IA, et contribuent ainsi à l'évolution des pratiques musicales avec le numérique. Nous avons donc ciblé deux communautés internationales habituées à discuter à propos des technologies employées ou employables en musique.

La première communauté de musiciens observée est la communauté en ligne Gearspace https://gearspace.com/board/ hébergée sur un site web de la société anglaise Gearspace limited et disposant d'un forum dédié à l'ingénierie audio. Selon l'université de Vermont (Vermont, 2021), ce forum créé en 2002 est l'une des plus grandes sources d'information sur l'audio professionnel, avec plus de 1,6 million de visiteurs mensuels provenant de 218 pays. La deuxième communauté de musiciens observée est la communauté KVR https://www.kvraudio.com/forum hébergée par la société américaine KVR audio. Cette communauté est constituée de plus de 600 000 membres provenant eux aussi des cinq continents. Les membres de ces communautés discutent dans des espaces thématiques différenciés tels que les dernières actualités sur les instruments, les effets, les logiciels utilisés en musique assistée par ordinateur, classés par catégories, les dernières annonces des fabricants, mais aussi sur les configurations matérielles, les productions réalisées, et d'autres thématiques générales que les petites annonces ou des sujets de réflexion transversaux à ces domaines.

Ces communautés sont sensibilisées à la thématique de l'intelligence artificielle en musique et se sont déjà largement exprimées sur cette thématique depuis plusieurs années avec des avis parfois très différents et tranchés. En observant ces deux communautés, nous cherchons à comprendre quelles sont les croyances et les représentations projetées par les participants sur l'usage de l'intelligence artificielle en musique et sur les risques perçus identifiés (Astier et Labour, 2021), depuis les médiations entre l'humain, la création et la machine, pour caractériser les nouveaux rapports, les nouvelles relations et collaborations qui se confrontent autour de ces technologies.

### 1.4. Données et méthodologie

Pour observer ces médiations et les processus d'appropriation de l'IA dans la création musicale, nous nous appuyons sur des communautés en ligne de musiciens que nous observons depuis notre cadre d'analyse. L'utilisation d'outils contrastifs comme TXT-M (Heiden *et al.,* 2010) nous permet de segmenter le corpus, de lemmatiser (Souvay et Pierrel, 2009), c'est-à-dire de « fournir, pour une forme donnée, la représentation standardisée du mot correspondant, utilisée le plus souvent en entrée dans un dictionnaire de référence » et d'utiliser la méthode dite des spécificités, selon le modèle hypergéométrique et l'observation des suremplois et des sous-emplois. Par ailleurs, nous nous appuyons sur la méthode *Alceste* (Leblanc, 2015), fondée sur une analyse statistique distributionnelle de type harrissien en référence aux travaux de Harris (1954 ; 1968 ; 1991), qui distingue les 25 textes qui relèvent de domaines spécialisés de ceux qui relèvent de la langue générale, pour mettre en évidence les grandes articulations du corpus, et les mondes lexicaux, en classant les énoncés du texte en fonction de la distribution de leur vocabulaire.

## 2. Représentations de l'intelligence artificielle en musique

### 2.1. L'IA créative : un perroquet stochastique ?

Dans la première partie de ce deuxième volet, nous revenons sur les usages des musiciens utilisateurs d'IA identifiés à partir des corpus étudiés. Nous présentons les résultats de l'analyse des usages des IA par les musiciens en contexte de création en observant notamment leurs motivations, leurs stratégies pour produire du matériau musical et l'évaluation qu'ils portent sur les résultats obtenus depuis celles-ci à partir des intelligences mobilisées. Pour ce faire, nous avons créé un corpus à partir de la collecte de l'ensemble des avis exprimés sur le sujet entre 2022 et 2024, correspondant à plus de 4244 contributions représentant plus de 2000 pages de texte. Les avis exprimés sont contrastés et pour la plupart argumentés même si des émotions sont aussi exprimées avec des émoticônes, émotions qui renvoient à l'ironie ou le scepticisme.

Les musiciens expriment diverses raisons d'utiliser l'IA dans la composition musicale. Le premier argument invoqué est celui de l'efficacité et de la rapidité de traduire l'idée musicale en geste musical automatisé avec des données Audio ou de type MIDI (Musical Instrument Digital Interface), comparativement aux méthodes utilisées dans le cadre de la musique assistée par ordinateur, sans l'usage d'intelligences artificielles. L'argument paraît essentiel pour les musiciens professionnels dans les contextes commerciaux où le temps est un facteur essentiel. La notion de rentabilité est récurrente dans ses applications pour générer des musiques d'illustration pour le montage vidéo ou la publicité, où, dans ce cas, la musique générée par l'IA paraît une option plus abordable que l'embauche de compositeurs humains.

Le deuxième argument évoqué dans les deux forums étudiés est celui de l'inspiration. Pour certains contributeurs, le recours à des intelligences artificielles permet d'obtenir des mélodies, des harmonies et des arrangements, sinon originaux, tout du moins efficaces, leur permettant d'améliorer leur processus de création et ce, quel que soit le niveau de formation ou de pratique du musicien, d'où la notion d'accessibilité associée à cet argument. Selon les membres de ces communautés, les outils d'IA permettraient de rendre la création musicale plus accessible à ceux qui n'ont pas de formation formelle ou à ceux qui n'ont pas la possibilité de collaborer ou qui ne souhaitent pas collaborer avec d'autres musiciens.

Le troisième argument émis est celui de l'exploration de nouveaux sons par la combinaison en couches (layers) de plusieurs synthèses, additives, granulaires, FM, ou à base d'échantillons, leur permettant d'explorer et de définir de nouveaux espaces sonores beaucoup plus facilement que par l'empilement de synthétiseurs, échantillonneurs ou expandeurs reliés en MIDI ou par la juxtaposition de plug-ins virtuels (VST ou AU) sur des pistes MIDI. De nouveaux styles et sons qu'ils n'auraient pas envisagés, élargissent ainsi leurs horizons créatifs. La rationalisation des processus de production musicale par l'automatisation de certaines tâches, telles que la génération d'échantillons audio ou la suggestion de matériel musical, sont aussi quelques-unes des finalités retenues par les musiciens utilisateurs, parfois producteurs, qui cherchent à augmenter le flux de travail et la productivité au cours des premières étapes de la création musicale (Bader, 2018 ; Verma, 2021).

Cependant, un nombre important de ces contributions renvoient au sentiment de déception quant à la qualité des résultats obtenus. Un tiers des contributeurs estiment que les résultats sont beaucoup trop aléatoires ou ne répondent pas à leurs attentes. L'idée défendue est que si l'usage de l'IA en musique permet de produire du matériau musical, elle ne parvient cependant pas, dans la majorité des cas, à générer et à apporter quelque chose de véritablement utile ou significatif à l'œuvre créée. Certains membres qualifient même les intelligences artificielles de « perroquets stochastiques », suggérant qu'elles se contentent de reproduire des modèles existants plutôt que de produire des idées originales. Dans l'ensemble, ces résultats suggèrent une relative transposition des profils des musiciens déjà utilisateurs de la musique assistée par ordinateur (MAO) sur ces intelligences artificielles, en soulignant la prédominance de la familiarité avec l'usage de logiciels et de plug-ins d'aide à la création sonore sur la prédisposition à utiliser ces intelligences artificielles.

### 2.2. Enjeux de la création sonore et musicale avec l'IA, depuis des interfaces. Du compagnon créatif au compagnon subversif

Dans la deuxième partie de ce deuxième volet, nous revenons sur les contributions émises par la deuxième communauté de musiciens observée, la communauté KVR https://www.kvraudio.com/forum hébergée par la société américaine KVR audio. Cette communauté est constituée de plus de 600 000 membres provenant eux aussi des cinq continents. Parmi les 512 membres qui ont consulté le sujet, 106 d'entre eux ont accepté de répondre à un sondage et 12 ont exprimé un avis sur l'usage de l'intelligence artificielle en musique. Nous complétons donc les contributions recueillies par ces avis exprimés autour de l'utilisation de l'intelligence artificielle dans la musique. Dans ce nouveau corpus, les contributeurs abordent le concept d'interface de plusieurs manières, en particulier dans le contexte de la création musicale et de l'ingénierie du son.

Le premier sentiment qui domine dans le sondage à l'égard de l'IA est le scepticisme chez 23 % des membres sondés qui ne voient pas l'intérêt de l'utiliser, préférant s'en remettre à leur propre créativité et à leurs propres processus. La notion de plaisir explique en partie ce désintérêt : « °Je ne trouve aucun plaisir à utiliser l'IA dans la musique ou les arts. C'est en fait un fardeau !° » D'autres y voient *a contrario* la possibilité de trouver de nouvelles idées et inspirations bien qu'ils reconnaissent que la qualité actuelle de la musique générée par l'IA n'est pas encore satisfaisante. Ils affirment cependant que si la technologie peut aider, elle ne doit pas remplacer l'élément humain dans la création musicale. « °Je ne suis donc pas opposé à l'intelligence artificielle pour les aspects techniques, dans les processus de mixage/mastering par exemple, mais je crée ma propre musique, d'où ma réponse : il n'y a pas de raison de l'utiliser° ».

D'autres considèrent l'IA comme un outil similaire à la technologie musicale traditionnelle (comme les pédales de guitare), suggérant qu'elle peut améliorer la créativité et l'efficacité sans diminuer la valeur de la musique créée par l'homme. Ils affirment que toute nouvelle technologie fait l'objet de critiques, mais qu'en fin de compte, l'art est subjectif.

Les avis sont partagés sur les considérations économiques, comme sur la question de savoir si l'IA peut produire des résultats à moindre coût. Si certains pensent que son usage pourrait réduire les coûts, d'autres affirment que la consommation d'énergie et de données requise par les outils d'IAGen pourrait annuler ces économies. En revanche, les avis se rejoignent pour considérer que les IAGen utilisées en musique permettent de produire des résultats plus rapidement, plus facilement, et de tester des idées musicales.

Au-delà du sondage et des commentaires exprimés autour de celui-ci, voire dans d'autres sujets connexes, plusieurs membres évoquent le risque d'obsolescence dans la représentation des structures musicales à l'écran à mesure que la technologie de l'intelligence artificielle progresse, et notamment dans le domaine de la création musicale réalisée depuis les interfaces des stations de travail audio-numériques (DAW). Par exemple, dans le cas de l'édition de partitions, ils suggèrent que les interfaces actuelles, qui obligent les utilisateurs à saisir manuellement des notes et des sons, peuvent être considérées comme *nostalgiques* plutôt que *pratiques*. Ces propos laissent à penser, de la part des contributeurs, des représentations des interfaces futures ne demandant finalement aux musiciens utilisateurs qu'un bagage musical minimum pour utiliser les intelligences créatives proposées et permettant de générer de la musique à partir de données simples, telles que des descriptions textuelles.

*Figure 1 : Question posée sur le forum international KVR*

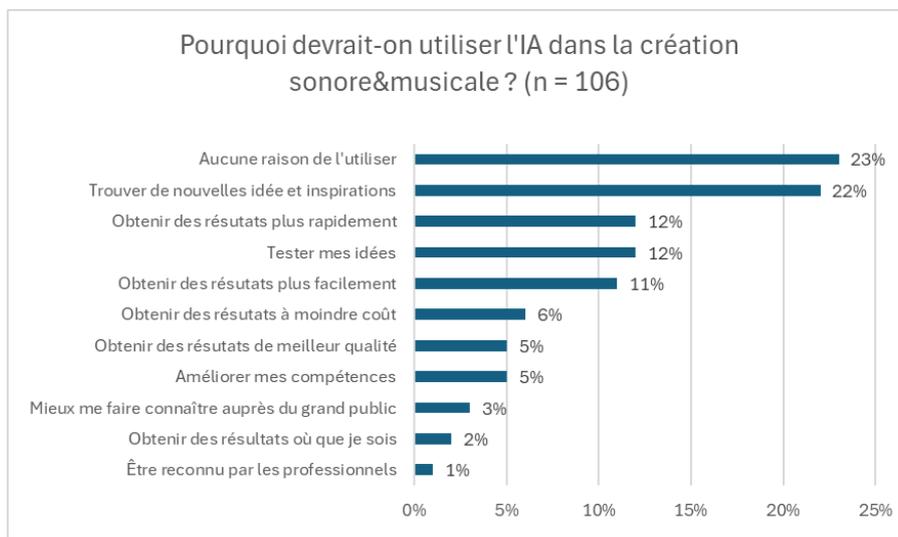

Dans le domaine du son, certaines contributions mentionnent la possibilité de concevoir des interfaces qui permettent des ajustements en temps réel de la sortie sonore, en mélangeant la musique générée par l'IA avec les signaux originaux, permettant ainsi une interaction en temps réel entre le musicien et la machine. L'interface est alors appréhendée comme un des composants du système qui donne priorité à l'interaction avec le musicien utilisateur et au retour d'information immédiat, permettant ainsi d'améliorer le processus créatif. Même si une partie de ces possibilités d'interaction en temps réel musicien-machine existent déjà depuis l'apparition de la norme MIDI et plus encore dans sa version 2, avec un passage de 128 à 16384 niveaux de contrôle comme dans des applications telles que Max for Live de Ableton (Rixte, 2024), les membres des communautés ont cependant évoqué la possibilité d'implémenter de nouvelles fonctions d'IA permettant d'augmenter le niveau et la qualité d'interaction par l'apprentissage par ces IA des gestes musicaux du musicien. D'où un questionnement possible sur le *sens* accordé de leur part, aux termes *niveau* et *qualité*. Nous suggérons, à l'instar des interfaces graphiques, que les attentes exprimées portent sur l'implémentation dans ces agents, de « propriétés adaptatives, issues des interactions », bien au-delà de « propriétés d'expressivité étendues ».

D'autres membres évoquent le concept Interface cerveau-parleur, le *Brain-to-Speaker Interface*, qui permettrait théoriquement à l'utilisateur de créer de la musique, en l'imaginant. Cependant, ils soulignent également les limites d'une telle technologie, en insistant sur le fait qu'une utilisation efficace nécessiterait une représentation mentale claire et structurée des éléments musicaux, représentation qui va à l'encontre de la part d'improvisation ou d'intuition présente dans le processus compositionnel non pas devant l'écran, mais sur un instrument connecté en MIDI à un séquenceur audionumérique tel qu'un clavier maître. La création et l'implémentation de ce concept soulèvent aussi, selon eux, la question de la relation complexe entre la cognition humaine et les interfaces technologiques qui, de compagnons créatifs se transformeraient en compagnons subversifs, voire en ennemis avec des risques de bugs et de manipulation de la part de la machine.

Cependant, la plupart des contributeurs appréhendent l'interface comme un pont entre la créativité humaine et les capacités de l'IA, permettant aux musiciens de saisir leurs idées et préférences musicales, que l'IA traite ensuite pour générer une musique conforme à leur vision créatrice. Cet aspect collaboratif est mis en avant dans les études qui explorent la manière dont les musiciens peuvent tirer parti des outils d'IA pour améliorer leurs flux de travail créatifs (Nicholls *et al.*, 2018). Dans celles-ci, l'interface est abordée comme un élément dynamique et évolutif de la technologie musicale, l'objectif étant de rendre la création musicale plus accessible et plus intuitive grâce aux progrès de l'IA et de la conception de l'expérience utilisateur. Une étude de Li *et al.,* (2025) a récemment exploré le paysage des méthodologies de décodage alimentées par l'IA pour les interfaces multimodales cerveau-ordinateur, en abordant leurs éléments, les algorithmes de décodage, les applications et les défis. En mettant l'accent sur les avancées en matière de cartographie multimodale, de modélisation séquentielle et de fusion multimodale, ce travail a mis en lumière le rôle essentiel des algorithmes de décodage de l'IA dans la mise au point d'interfaces multimodales encore plus précises. D'autres recherches (Postolache *et al.*, 2025) continuent d'explorer le potentiel des modèles de diffusion latente, une famille de modèles génératifs puissants, à base d'intelligence artificielle pour la reconstruction de musique naturaliste à partir d'enregistrements d'électro-encéphalogrammes (EEG). Contrairement à la musique générées par MIDI, l'accent est mis sur la génération de musique plus complexe comportant un éventail varié d'instruments, de voix et d'effets, riche en harmoniques et en timbres. Les premiers résultats obtenus montrent la nécessité de recherches supplémentaires pour améliorer la généralisation du changement de distribution, en termes d'ensembles de données produites à partir d'algorithmes devant être améliorés. Dans une revue systématique de littérature, Sayal *et al.,* (2025) explorent l'utilisation de la musique pour permettre l'autocontrôle de l'activité cérébrale et ses implications en neurosciences cliniques. La revue suggère de prendre en compte les corrélats neuronaux du cerveau entier, les stimuli musicaux et leur interaction, avec les réseaux cérébraux cibles et les mécanismes de récompense lors de la conception d'études de neurofeedback musical. Les auteurs suggèrent, en ce qui concerne la spécificité de l'interface musicale et ses avantages potentiels en termes de récompense, qu'aucune des études analysées n'a comparé différentes interfaces de rétroaction et que de ce fait, aucune conclusion ne peut être tirée à ce stade, concernant la causalité entre le succès et l'interface musicale.

*A contrario*, plusieurs études (Civit *et al.*, 2022 ; Li & Wang, 2024) ont montré comment la conception et la fonctionnalité de ces interfaces peuvent améliorer ou entraver l'expérience du musicien-utilisateur dans la création ou l'enseignement de la musique, selon le niveau d'accessibilité permettant à ces systèmes d'être utiles et utilisables (Tricot, 2020). Cette accessibilité est cruciale pour favoriser la créativité et encourager l'expérimentation de la musique générée par l'IA (Li & Wang, 2024). Ainsi, l'intégration d'éléments visuels tels que des représentations graphiques du son et des commandes interactives, aide les utilisateurs à comprendre les processus et les résultats de l'IA, ce qui rend la technologie plus accessible (Civit *et al.*, 2022).

En résumé, pour les contributeurs, la conception de l'interface et sa qualité ergonomique (Scapin et Bastien, 1995) peut avoir un impact significatif sur la manière dont les utilisateurs perçoivent le rôle de l'IA dans le processus créatif, influençant leur volonté d'adopter et d'expérimenter ces technologies (Zhang, 2020). L'interface de l'intelligence artificielle (IA) dans la musique est donc considérée comme un élément essentiel qui influence l'interaction avec l'utilisateur, la créativité et l'efficacité globale des outils d'IA.

## 3. Discussion : appropriation et confrontation culturelle aux intelligences artificielles, une collaboration créative encore individualiste

Dans l'analyse des contributions, c'est la possibilité par un humain de demander à l'IA, sans limites, des changements spécifiques ou de nouvelles idées, qui suscite l'intérêt et favorise une interaction dynamique avec l'IA dans l'espace de travail, susceptible de conduire à des résultats musicaux inattendus et innovants.

Pour autant, les processus d'appropriation de ces espaces (Lai *et al.*, 2024) génèrent de grandes inquiétudes car si la possibilité de dialogue entre l'humain et la machine est perçue positivement puisque permettant de modifier et d'affiner les résultats produits par l'IA pour les aligner sur la vision artistique du musicien utilisateur, des risques de confrontation ou de surenchère entre les parties prenantes avec le risque de perdre le fil conducteur de la pensée musicale initiale, sont redoutés. La crainte d'une appropriation inachevée de l'espace le conduisant à une forme de dépendance à l'IA qui risque d'étouffer la créativité et les compétences au fil du temps, est particulièrement appréhendée. Les musiciens craignent en ce sens une sorte d'homogénéisation dans les résultats des médiations humain-machine, l'homogénéisation pouvant par exemple provenir des limites de la norme MIDI au niveau de l'expressivité et des micro-tonalités, de l'utilisation d'algorithmes standardisés (Ahuja, 2024) reproduisant les modèles de réussite, ou encore de prompts identiques conduisant à des résultats produits par l'IA, très proches les uns des autres.

D'autres contributeurs craignent une perte de l'interaction humaine dans le processus de création, interaction qui apparaît pourtant comme essentielle à la créativité. Ils redoutent aussi que la qualité de la musique générée par l'IA ne soit pas à la hauteur de la profondeur et de la richesse des compositions créées par des musiciens humains, avec le risque que les aspects émotionnels et expérientiels issus de de la collaboration humaine soient perdus par la substitution avec une IA. Des chercheurs (Mitra et Zualkerman, 2025) ont cependant montré que les technologues accordent de l'importance aux erreurs et aux bogues dans le processus avec la machine, la valeur accordée aux erreurs démontrant leur association avec des valeurs émotionnelles.

Pour autant, les collaborations entre les humains et l'IA sont tout de même considérées comme un moyen de surmonter des blocages créatifs. Ces blocages ont été repérés dans la littérature (Kumar *et al.*, 2025) où quatre obstacles majeurs à la créativité ont été explicités (Kreminski *et al.*, 2019), à savoir, la peur de la toile blanche ; la peur du jugement ; la peur de l'échec ; et le perfectionnisme. Ils rejoignent en ce sens les travaux qui suggèrent qu'en utilisant des instructions générées par l'IA, les musiciens peuvent franchir ces obstacles et explorer de nouvelles directions dans leur travail (Li & Wang, 2024), en questionnant d'abord leur motivation à créer avec l'IA (Chen, 2025) ou en utilisant trois types d'outils : (1) les réseaux neuronaux (NN) comme WaveNet et TimbreTRon pour la reconstruction audio et la transformation du timbre ; (2) les auto-codeurs variationnels (VAE) comme MIDI-VAE pour analyser et produire des informations sur la dynamique des hauteurs et sur les performances des instruments dans la musique polyphonique ; (3) ou encore les transformateurs comme MusIAC pour accroître l'évolutivité et la contrôlabilité des structures musicales et améliorer l'intégrité structurelle et l'expressivité de la musique (Yang *et al.,* 2024).

Par ailleurs et du point de vue des dimensions psychosociales de l'appropriation, certains musiciens redoutent des usages déviants de ces espaces de travail par rapport a ce qui a aurait été prévu ou anticipé par les concepteurs (Cobaleda et Babapour, 2017 ; Vischer, 2008), non seulement de la part de l'utilisateur mais aussi de l'IA avec, dans les collaborations possibles, des risques d'asservissement à l'IA :

« Supposons que je dépende de l'IA Drums dans Logic, même si j'ai toutes les commandes, je serai "esclave" du résultat produit par l'IA ».

S'agissant de la facilité d'appropriation (Nielsen, 1994), la collaboration entre les humains et l'IA dans le processus de création musicale est aussi appréhendée comme une boucle de rétroaction, dans laquelle les musiciens guident la production en interagissant avec le contenu généré, avec la possibilité d'annuler ou de refaire, d'où, une certaine facilité de se souvenir (Nielsen, 1994). La conception issue de ce type de collaboration possible rejoint le point de vue de certains auteurs (Yang *et al.*, 2024) qui voient dans le processus itératif, la possibilité d'améliorer la qualité de la musique produite, monodique ou polyphonique, en raison de la capacité des réseaux neuronaux à gérer des dépendances à long terme et des mécanismes d'interconnexions, permettant par exemple le respect de règles d'écriture dans la proposition artificiellement générée, selon le style musical ou l'esthétique recherchée. D'autres auteurs (Ben-Tal *et al.,* 2021) suggèrent la possibilité pour les musiciens de conserver le contrôle créatif de l'idée et du geste musical, tout en bénéficiant des capacités génératives de l'IA, à savoir, la création de mélodies, d'accords, et plus largement le traitement sans limites de structures temporelles de données musicales permettant ainsi d'améliorer l'intégrité structurelle et l'expressivité de la musique (Yang *et al.,* 2024).

Cette dernière possibilité idéalise les collaborations possibles finalement dans lesquelles l'IA serait reléguée à l'exécution de certaines tâches répétitives de paramétrage ou génératives de contenu aléatoire à l'infini, ou au contraire de tâches normalisées selon des règles établies comme l'orchestration, tâches considérées comme mécaniques et de moindre intérêt que les tâches purement créatives, où l'inspiration humaine est encore considérée comme un idéal. C'est en réponse à cette vision qui avait déjà été repérée, que des chercheurs travaillent sur la conception de systèmes comme « Stable Audio » et « Magenta Studio », qui permettent aux musiciens de saisir des paramètres ou des thèmes spécifiques, depuis lesquels l'IA génère un contenu musical de base propice à des développements ultérieurs (Civit *et al.*, 2022 ; Nicholls *et al.*, 2018). Les nouvelles formes de collaboration entre les humains et l'IA dans le processus de création musicale sont là encore appréhendées comme une relation synergique individuelle susceptible d'améliorer la créativité.

En résumé, il ressort de l'analyse de ces contributions mises en parallèle avec la littérature, une vision encore individualiste du musicien dans son rapport à la créativité collaborative avec l'IA, alors même qu'une des caractéristiques de cette collaboration créative est que « la nature et la qualité des interactions entre les membres de l'ensemble sont un déterminant essentiel des résultats musicaux » (Hill et Fitzgerald, 2012). Si les collaborations complémentaires (John-Steiner, 2000), qui reposent sur la reconnaissance et l'exploitation d'expertises, de connaissances disciplinaires, de rôles et de tempéraments complémentaires pour poursuivre un objectif commun, semblent présentes dans les communautés où les musiciens s'interrogent, elles ne sont pas exprimées dans une perspective socio-constructiviste par les membres au sujet de la créativité et des processus d'appropriation des IA.

**Conclusion**

Nous avons présenté deux communautés numériques de musiciens, et avons observé depuis leurs contributions, leurs approches des activités de création, d'interprétation, d'édition et de diffusion avec l'intelligence artificielle (IA) créative. Notre analyse permet d'identifier un rapport d'attirance et de rejet de ces intelligences intégrées dans la production musicale en raison de leurs impacts sur la créativité, l'authenticité, qui remodèlent le rapport musicien-machine. Ce rapport peut notamment s'observer depuis de nouvelles interfaces elles aussi dotées d'intelligences artificielles et qui nécessitent de nouvelles appropriations avant de permettre de nouvelles médiations, constituant ainsi de nouveaux enjeux pour les musiciens. D'où, selon le niveau d'appropriation où ils se trouvent, une perception de l'IA et de sa place dans l'acte de faire de la musique (*musing doing*), plus ou moins tranchée et qui varie considérablement d'une personne à l'autre. Alors que certains membres semblent s'engager positivement dans le recours à l'IA en complément du geste musical, d'autres expriment leur scepticisme quant à l'authenticité et à la résonance émotionnelle du matériau musical produit. Il faut peut-être y voir l'émergence de nouveaux paradigmes de création et d'interprétation dans un paysage actuel qui pourrait évoluer vers un modèle de collaboration dans lequel l'IA servirait d'assistant aux compositeurs ou donnerait la réplique aux interprètes humains. Ce partenariat pourrait permettre aux musiciens d'exploiter les suggestions générées par l'IA tout en gardant le contrôle de la création, favorisant ainsi des approches novatrices de la création musicale. De tels développements pourraient conduire à une musique adaptative en temps réel, personnalisée en fonction des préférences de l'auditeur, brouillant encore davantage les frontières entre l'art humain et l'art généré par la machine. Cet aspect de collaboration met en évidence le potentiel de l'IA à agir comme une source d'inspiration et un catalyseur pour l'exploration créative, qui suscite toutefois de nombreuses craintes comme le risque d'uniformisation des contenus, la perte d'activité, la concurrence voire la dépendance.

Si la collaboration entre les humains et l'IA dans la création musicale est perçue comme une offre de possibilités de collaborations, elle est également appréhendée comme un ensemble de technologies porteuses de risques importants liés à la créativité, au développement des compétences, aux considérations éthiques et aux difficultés à s'approprier les médiations qui se déroulent dans des espaces de travail interconnectés où des collaborations multidimensionnelles humain-machine restent encore à appréhender.

**Bibliographie**

Ahuja, A. (2024). Melodic Deception: Exploring the Complexities of Deepfakes of Music Generated by Generative Adversarial Networks (GANs*). Sangeet Galaxy, 13*(2).

Astier, B. et Labour, M. (2021). Walk the talk – risque perçu et innovation collaborative : le cas de Plateau Fertile, un tiers-lieu textile. *Approches Théoriques en Information-Communication (ATIC)* N° 2(1), 47-73. https://doi.org/10.3917/atic.002.0047.

Ashri, R. (2019). *The AI-powered workplace: how artificial intelligence, data, and messaging platforms are defining the future of work.* Apress.

Bachimont, B., & Crozat, S. (2004). Instrumentation numérique des documents : pour une séparation fonds/forme. *Revue I3-Information Interaction Intelligence*, *4*(1).

Bader, R. (2018). The relation between music technology and music industry. In R. Bader (Ed.), *Springer Handbook of Systematic Musicology* (pp. 899-901). Springer-Verlag GmbH Germany.

Bamigbola, A. (2021). Artificial intelligence in music education: A review. *International Journal of Music Education*, 39(3), 345-358.

Barrett, M.S. (Ed.). (2014). *Collaborative creative thought and practice in music*. Farnham: Ashgate Publishing Limited.

Barrett, M. S., Creech, A., & Zhukov, K. (2021). Creative collaboration and collaborative creativity: A systematic literature review. *Frontiers in Psychology, 12*, 713445.

Beenkens, F.H.C. and Verburg, R.M. (2007), "Extending TAM to measure the adoption of E-collaboration in healthcare arenas", in Kock, N. (Ed.), Encyclopedia of E-Collaboration, IGI Global, Hershey, PA, pp. 265-272.

Ben-Tal, O., Harris, M. T., & Sturm, B. L. (2021). How music AI is useful: Engagements with composers, performers and audiences. *Leonardo*, 54(5), 510–516. https://doi.org/10.1162/leon_a_01959

Blackwell, A. F., Damena, A., & Tegegne, T. (2021). Inventing artificial intelligence in Ethiopia. *Interdisciplinary Science Reviews*, 46(3), 363–385. https://doi.org/10.1080/03080188.2020.1830234

Briot J.-P. (2021). From artificial neural networks to deep learning for music generation: history, concepts and trends. *Neural Computing and Applications, 33* (1), p. 39-65.

Chamberlain, A., Hazzard, A., Kelly, E., Bødker, M., & Kallionpää, M. (2021). From AI, creativity and music to IoT, HCI, musical instrument design and audio interaction: A journey in sound. *Personal and Ubiquitous Computing*, 25(4), 617–620. https://doi.org/10.1007/s00779-021-01554-z

Civit, M., Civit-Masot, J. Cuadrado, F. et Escalona, M. J. (2022). A systematic review of artificial intelligence-based music generation: Scope, applications, and future trends, *Expert Systems with Applications, Volume 209*, 118190. https://doi.org/10.1016/j.eswa.2022.118190. (https://www.sciencedirect.com/science/article/pii/S0957417422013537 )

Cobaleda Cordero, A., & Babapour Chafi, M. (2017). Discrepancies between intended and actual use in Activity-based Flexible Offices—A literature review. 11. In *Joy at Work'Nordic Ergonomics Society Conference Proceedings*, *NES* (p. 56).

Corazza, G. E. (2019a). *"The dynamic universal creativity process" in Dynamic perspectives on creativity: New directions for theory, research, and practice in education*. eds. R. A. Beghetto and G. E. Corazza (Cham: Springer), 297–318.

Corazza, G. E., & Lubart, T. (2020). The big bang of originality and effectiveness: A dynamic creativity framework and its application to scientific missions. *Frontiers in Psychology*, *11*, 575067.

Dan, L., Naiyao, Z., & Hancheng, Z. (2003). A CAD system of music animation based on form and mood recognition. *Pattern Recognition and Artificial Intelligence*, *16*(3), 283–287.

Dash, A., & Agres, K. (2024). Ai-based affective music generation systems: A review of methods and challenges. *ACM Computing Surveys, 56*(11), 1-34.

Dey, B.L., Binsardi, B., Prendergast, R. and Saren, M. (2013), "A qualitative enquiry into the appropriation of mobile telephony at the bottom of the pyramid", *International Marketing Review, 30* (4), pp. 297-322.

Ernst, E., Merola, R., & Samaan, D. (2019). Economics of artificial intelligence: Implications for the future of work. *IZA Journal of Labor Policy*, 9(1), 1–35. https://doi.org/10.2478/izajolp-2019-0004

Garnham, N. (1999), "Amartya Sen's 'Capability' approach to the evaluation of welfare: its application to communications", in *Calabrese, A. and Burgelman, J.-C. (Eds), Communication, Citizenship, and Social Policy: Rethinking the Limits of the Welfare State*, Rowman & Littlefield Publishers, Lanham, MD

Glăveanu, V. P., Gillespie, A., & Valsiner, J. (2015). *Rethinking creativity*. East Sussex: Routledge.

Glăveanu, V.P., Hanson, M.H., Baer, J., Barbot, B., Clapp, E.P., Hennessey, B., Kaufman, J.C., Lebuda, I., Lubart, T., Montuori, A., Ness, I.J., Plucker, J., Reiter-Palmon, R., Sierra, Z., Simonton, D.K., Neves Pereira, M.S., & Sternberg, R.J. (2019). Advancing creativity theory and research: A socio-cultural manifesto. *Journal of Creative Beahviour*. Available at: https://onlinelibrary.wiley.com/doi/full/10.1002/jocb.395?fbclid=IwAR1OpJ2bmqneyQJECMchh7OpBHGRhg6e0ueTDZIz7mdXJHZ470xStsxpJUU

Gouzouasis, P., & Bakan, D. (2011). The future of music making and music education in a transformative digital world. *The University of Melbourne Refereed E-Journal*, *2*(2), 127–154.

Harris, Z. (1954). Distributional structure. *Word, 10*(23):146–162. 278

Harris, Z. (1968). *Mathematical structures of language*. John Wiley & Sons.

Harris, Z. (1991). *A theory of language and information: a mathematical approach*. Clarendon Press; Oxford University Press, Oxford, England.

Heiden, S., Magué, J. P., & Pincemin, B. (2010, June). TXM : Une plateforme logicielle open-source pour la textométrie-conception et développement. In *10th International Conference on the Statistical Analysis of Textual Data-JADT 2010; 2*(3), pp. 1021-1032). Edizioni Universitarie di Lettere Economia Diritto.

Hennion A, (1993a). *La passion musicale : une sociologie de la médiation*. Paris : Edition Métailié.

Hennion, A. (1993b). L'histoire de l'art : leçons sur la médiation. *Réseaux. Communication-Technologie-Société, 11*(60), 9-38.

Hill, B., and Fitzgerald, J. (2012). Human machine music: an analysis of creative practices among Australian live electronica musicians. *Perfect Beat 13*, 161–180. doi: 10.1558/prbt.v13i2.161

Janneck, M. (2009), "Recontextualising technology in appropriation Processes", in *Whitworth, B. and Moor, A.d. (Eds), Handbook of Research on Socio-Technical Design and Social Networking Systems*,

Jeanneret, Y., & Ollivier, B. (2004). *Les sciences de l'information et de la communication : savoirs et pouvoirs, 38*. CNRS.

John-Steiner, V. (2000). *Creative collaboration*. New York: Oxford University Press.

Kellner, C., Massou L. et Morelli P. (2010). (Re)penser le non-usage des tic. *Questions de communication, 18,* 7-20, https://doi.org/10.4000/questionsdecommunication.395.

Kenny, A. (2014). Collaborative creativity'within a jazz ensemble as a musical and social practice. *Thinking Skills and Creativity, 13*, 1-8.

Kirk, C.P., Swain, S.D. and Gaskin, J.E. (2015), "I'm proud of it: consumer technology appropriation and psychological ownership", *Journal of Marketing Theory and Practice, 23* (2), pp. 166-184.

Lai, C., Ianeva, M., Bobillier Chaumon, M. É., & Abitan, A. (2021). Une perspective située pour penser l'appropriation des espaces de travail « par activités ». *Activités*, (18-2).

Leblanc, J. M. (2015). Proposition de protocole pour l'analyse des données textuelles : pour une démarche expérimentale en lexicométrie. *Nouvelles perspectives en sciences sociales, 11*(1), 25-63.

Li, P., & Wang, B. (2022). Artificial intelligence in music education: A study on the effectiveness of AI-enabled teaching tools. *Journal of Music Technology and Education*, *15*(2), 123-138.

Li, S., Wang, H., Chen, X., & Wu, D. (2025). Multimodal Brain-Computer Interfaces: AI-powered Decoding Methodologies. *arXiv preprint arXiv:2502.02830*.

Lorrain, D. (2003). Réalisation de jeux musicaux du XVIIIe siècle : Mozart & Stadler. In *Journées d'Informatique Musicale*.

Makridakis, S. (2017). The forthcoming Artificial Intelligence (AI) revolution: Its impact on society and firms. *Futures*, *90*, 46-60.

Mora-Gutiérrez Paitan , F. M., Meléndez, M. A., Ovalle, C. (2024). Application of artificial intelligence in music generation: a systematic review. *IAES International Journal of Artificial Intelligence 13 (4)*, 3715-3726.

Newman, M., Morris, L., & Lee, J. H. (2023, November). Human-AI Music Creation: Understanding the Perceptions and Experiences of Music Creators for Ethical and Productive Collaboration. In *ISMIR* (pp. 80-88).

Nielsen, J. (1994). Estimating the number of subjects needed for a thinking aloud test. *International Journal of Human-Computer Studies, 41* (3), 385-397.

O'Brien, H.L. and Toms, E.G. (2008), "What is user engagement? A conceptual framework for defining user engagement with technology", in *Centre for Management Informatics*, D.U., UBC Faculty Research and Publications. 6100 University Avenue, Library, Halifax, Nova Scotia.

Postolache, E., Polouliakh, N., Kitano, H., Connelly, A., Rodolà, E., Cosmo, L., & Akama, T. (2025, April). Naturalistic music decoding from EEG data via latent diffusion models. In *ICASSP 2025-2025 IEEE International Conference on Acoustics, Speech and Signal Processing (ICASSP)* (pp. 1-5). IEEE

Rabardel, P. (1995). *Les hommes et les technologies ; approche cognitive des instruments contemporains* (p. 239). Armand colin.

Rixte, A. (2024). LiveScaler: Live control of the harmony of an electronic music track. *arXiv preprint arXiv:2401.08181*.

Runco, M. A., & Jaeger, G. J. (2012). The standard definition of creativity. *Creativity research journal*, *24*(1), 92-96.

Russell, S. J., & Norvig, P. (2020). *Artificial Intelligence: A Modern Approach* (3rd ed.). Pearson.

Sayal, A., Direito, B., Sousa, T., Singer, N., & Castelo-Branco, M. (2025). Music in the loop: a systematic review of current neurofeedback methodologies using music. *Frontiers in Neuroscience, 19*, 1515377.

Scapin, D. L., & Bastien, J. C. (1997). Ergonomic criteria for evaluating the ergonomic quality of interactive systems. *Behaviour & information technology*, *16*(4-5), 220-231.

Schiavio, A., van der Schyff, D., Biasutti, M., Moran, N., & Parncutt, R. (2019a). Instrumental Technique, Expressivity, and Communication. A qualitative study on learning music in individual and collective settings. Frontiers in *Psychology, 10*, 737.

Selwyn, N. (2004), "Reconsidering political and popular understandings of the digital divide", *New Media and Society, 6* (3), pp. 341-362.

Souvay, G., & Pierrel, J. M. (2009). LGeRM Lemmatisation des mots en moyen français. *Revue TAL : traitement automatique des langues, 50*(2), 21.

Tegmark, M. (2017). *Life 3.0 : Being Human in the Age of Artificial Intelligence*. Allen Lane.

Tricot, A., & Tricot, M. (2000, October). Un cadre formel pour interpréter les liens entre utilisabilité et utilité des systèmes d'information. In *Ergo-IHM 2000*. Association Francophone d'Interaction Homme-Machine.

Vear, C., Benford, S., Avila, J. M., & Moroz, S. (2023). Human-AI Musicking: A Framework for Designing AI for Music Co-creativity. *AIMC 2023*.

Verma, S. (2021). Artificial intelligence and music: History and the future perspective. *International Journal of Applied Research*, 7(2), 123-130.

Vischer, J. C. (2008). Towards an environmental psychology of workspace: how people are affected by environments for work. *Architectural science review, 51*(2), 97-108.

Wang, X., & Li, P. (2024). Assessment of the Relationship Between Music Students' Self-Efficacy, Academic Performance and Their Artificial Intelligence Readiness. *European Journal of Education, 59*(4), e12761.